\documentclass[fleqn,twoside]{article}
\usepackage{espcrc2}

\usepackage{graphicx}
\usepackage[figuresright]{rotating}

\newcommand{\AmS}{{\protect\the\textfont2
  A\kern-.1667em\lower.5ex\hbox{M}\kern-.125emS}}

\textfloatsep=\floatsep
\intextsep=\floatsep

\title{Solving the Schwinger-Dyson Equations for Gluodynamics 
       in the Maximal Abelian Gauge%
       \thanks{Supported by Sumitomo Foundations,
       Grant-in-Aid for Scientific Research  (B)13135203 from MEXT,
       and (C)14540243  from JSPS.}}

\author{T.~Shinohara%
        \address[GS]{Graduate School of Science and Technology, 
                       Chiba University, Chiba 263-8522, Japan}%
        \thanks{Speaker at the conference.},
        K.-I.~Kondo%
        \addressmark[GS]%
        \address{Department of Physics, Faculty of Science, 
                 Chiba University,  Chiba 263-8522, Japan}
        and
        T.~Murakami\addressmark[GS]}

\begin{document}

%##########%
% Abstract %
%##########%
\begin{abstract}
We derive the Schwinger-Dyson equations
for the $SU(2)$ Yang-Mills theory in the maximal Abelian gauge
and solve them in the infrared asymptotic region.
We find that the infrared asymptotic solutions for the gluon
and ghost propagators are consistent with the hypothesis
of Abelian dominance.
\vspace{1pc}
\end{abstract}

\maketitle

%%%%%%%%%%%%%%%%%%%%%%%%
\section{Introduction} %
%%%%%%%%%%%%%%%%%%%%%%%%
The Schwinger-Dyson (SD) equation
is one of the most popular approaches to investigate
the non-perturbative features of quantum field theory.
The analyses by making use of the SD equation
for quark propagator are well-known.
Recently, the coupled SD equations
for the gluon and ghost propagators
in Yang-Mills theory have been studied
mainly in the Lorentz (Landau) gauge.\cite{SAH,AB98}
In this paper, we derive the SD equations
for the $SU(2)$ Yang-Mills theory
in the maximal Abelian (MA) gauge
and solve them analytically
in the infrared (IR) asymptotic region.
The MA gauge is useful to investigate the Yang-Mills theory
from the view point of the dual superconductivity.

In the MA gauge,
in contrast to the ordinary Lorentz gauge,
we must explicitly distinguish
the diagonal components of the fields from the off-diagonal components.
This is indeed the case
even in the  perturbative analysis in the UV region.\cite{SIK03}
Therefore, we must take account of the four propagators
for the diagonal gluon, off-diagonal gluon,
diagonal ghost and off-diagonal ghost.
Numerical behaviors of gluon propagators in the MA gauge are also
investigated on a lattice simulation.\cite{BCGMP03}

%%%%%%%%%%%%%%%%%%%%%%%%%%%%%%%%%%%%%%%%
\section{SD equations in the MA gauge} %
%%%%%%%%%%%%%%%%%%%%%%%%%%%%%%%%%%%%%%%%
First, we derive the SD equations
from the $SU(2)$ Yang-Mills action
in the MA gauge\cite{SIK03}.
The graphical representation of SD equations are shown
in Figure~\ref{fig:SDE}.
%%%%%%%%%%%%%%%%%
% FIGURE(BEGIN) %
%%%%%%%%%%%%%%%%%
\unitlength=.001in
\begin{figure*}[tb]
\begin{picture}(6000,1800)
%\put(0,2000){\tframe[1000][100](6000,2000)}%
\put(0,-200){%
\put(0,500){%
\put(0,150){\includegraphics[height=.15in]{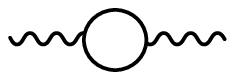}}%
\put(450,300){\mbox{$-1$}}%
\put(600,160){\mbox{\Large$=$}}%
\put(800,200){\includegraphics[height=.04in]{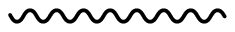}}%
\put(1250,300){\mbox{$-1$}}%
\put(1400,160){\mbox{\Large$+$}}%
\put(1600,0){\includegraphics[height=.4in]{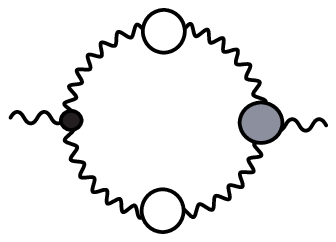}}%
\put(2000,350){\mbox{$3$}}%
\put(2200,160){\mbox{\Large$+$}}%
\put(2400,160){\mbox{\large(tadpole graphs)}}%
\put(3600,160){\mbox{\Large$+$}}%
\put(3800,160){\mbox{\large(two-loop graphs)}}%
}%
\put(0,1000){%
\put(0,150){\includegraphics[height=.16in]{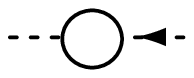}}%
\put(450,300){\mbox{$-1$}}%
\put(600,160){\mbox{\Large$=$}}%
\put(800,200){\includegraphics[height=.05in]{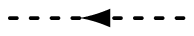}}%
\put(1250,300){\mbox{$-1$}}%
\put(1400,160){\mbox{\Large$+$}}%
\put(1600,100){\includegraphics[height=.3in]{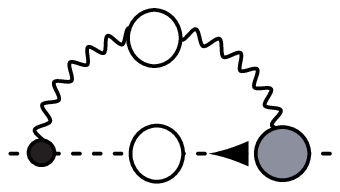}}%
\put(2000,350){\mbox{$3$}}%
\put(2200,160){\mbox{\Large$+$}}%
\put(2400,160){\mbox{\large(tadpole graphs)}}%
\put(3600,160){\mbox{\Large$+$}}%
\put(3800,160){\mbox{\large(two-loop graphs)}}%
}%
\put(0,1500){%
\put(0,150){\includegraphics[height=.15in]{Df.eps}}%
\put(0,250){\mbox{$3$}}%
\put(450,300){\mbox{$-1$}}%
\put(600,160){\mbox{\Large$=$}}%
\put(800,200){\includegraphics[height=.04in]{D0.eps}}%
\put(1000,250){\mbox{$3$}}%
\put(1250,300){\mbox{$-1$}}%
\put(1400,160){\mbox{\Large$+$}}%
\put(1600,0){\includegraphics[height=.4in]{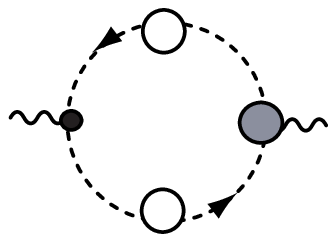}}%
\put(1570,230){\mbox{$3$}}%
\put(2200,160){\mbox{\Large$+$}}%
\put(2400,0){\includegraphics[height=.4in]{D1.eps}}%
\put(2370,230){\mbox{$3$}}%
\put(3000,160){\mbox{\Large$+$}}%
\put(3200,160){\mbox{\large(tadpole graphs)}}%
\put(4400,160){\mbox{\Large$+$}}%
\put(4600,160){\mbox{\large(two-loop graphs)}}%
}%
\put(0,0){%
\put(0,150){\includegraphics[height=.16in]{Gf.eps}}%
\put(0,250){\mbox{$3$}}%
\put(450,300){\mbox{$-1$}}%
\put(600,160){\mbox{\Large$=$}}%
\put(800,200){\includegraphics[height=.05in]{G0.eps}}%
\put(1000,250){\mbox{$3$}}%
\put(1250,300){\mbox{$-1$}}%
}%
}%
\end{picture}
\caption{The graphical representation of the SD equations.}
\label{fig:SDE}
\end{figure*}
%%%%%%%%%%%%%%%
% FIGURE(END) %
%%%%%%%%%%%%%%%
For the diagonal gluon propagator,
we adopt the Landau gauge so that the diagonal gluon propagator
$D_{\mu\nu}(p^2)$
has only the transverse part
\begin{equation}
\textstyle
D_{\mu\nu}(p^2)
 :=\frac{F_{\rm d}(p^2)}{p^2}
   P_{\mu\nu}^{\rm T},
\end{equation}
where we defined the form factor $F_{\rm d}(p^2)$.
While, the off-diagonal gluon propagator $D_{\mu\nu}^{ab}(p^2)$
has both the transverse and longitudinal parts
\begin{equation}
\textstyle
D_{\mu\nu}^{ab}
  :=\left[
    \frac{F_{\rm T}(p^2)}{p^2}
    P_{\mu\nu}^{\rm T}
%    \left(g_{\mu\nu}-\frac{p_\mu p_\nu}{p^2}\right)
    +\alpha\frac{F_{\rm L}(p^2)}{p^2}
    P_{\mu\nu}^{\rm L}
%    \frac{p_\mu p_\nu}{p^2}
    \right]\delta^{ab},
\end{equation}
where we defined the form factors $F_{\rm T}(p^2)$ and $F_{\rm L}(p^2)$.
The form factor $G(p^2)$ for the off-diagonal ghost propagator
${\mit\Delta}^{ab}(p^2)$ is defined
\begin{equation}
\textstyle
{\mit\Delta}^{ab}(p^2)
 :=\frac{G(p^2)}{p^2}\delta^{ab}.
\end{equation}
The diagonal ghost propagator is decoupled from the other fields
so that we omit it hereafter.

Now, we write down the SD equations:
\begin{eqnarray}
&&\!\!\!\!\!\!\!\!\!\!\!\!
D_{\mu\nu}(q^2)^{-1}
 =Z_aD_{\mu\nu}^{(0)}(q^2)^{-1}
    \nonumber\\
&&\!\!\!\!\!\!\!\!\!
%   \textstyle
   +Z_C^2%P_{\mu\nu}^{\zeta=4}
     \int_p
     %\int\frac{d^4p}{(2\pi)^4}
     {\mit\Gamma}_\mu^{(0)da}(r,p)
     {\mit\Delta}^{ab}(p)
     %\frac{G^{ab}(p)}{p^2}
     {\mit\Gamma}_\nu^{bc}(p,r)
     {\mit\Delta}^{cd}(r)
     %\frac{G^{cd}(r)}{r^2}
    \nonumber\\
&&\!\!\!\!\!\!\!\!\!
%   \textstyle
   -\frac{Z_A^2}2%P_{\mu\nu}^{\zeta=4}
     \int_p
     %\int\frac{d^4p}{(2\pi)^4}
     {\mit\Gamma}_{\mu\rho\xi}^{(0)da}(r,p)
     D_{\xi\eta}^{ab}(p^2)
     %\frac{G^{ab}(p)}{p^2}
     {\mit\Gamma}_{\nu\sigma\eta}^{bc}(p,r)
     %\frac{G^{cd}(r)}{r^2}
     D_{\rho\sigma}^{cd}(p^2)
    \nonumber\\
&&\!\!\!\!\!\!\!\!\!
  +\mbox{(tadpole contributions)},
\label{eq:diagonal gluon}
\end{eqnarray}
\begin{eqnarray}
&&\!\!\!\!\!\!\!\!\!\!\!\!
{\mit\Delta}^{ab}(q^2)^{-1}
 =Z_C{\mit\Delta}^{(0)ab}(q^2)^{-1}
    \nonumber\\
&&\!\!\!\!\!\!\!\!\!
   -Z_C^2
     \int_p
     {\mit\Gamma}_\mu^{(0)ac}(r,p)
     {\mit\Delta}^{cd}(p)
     {\mit\Gamma}_\nu^{db}(p,r)
     D_{\mu\nu}(r)
    \nonumber\\
&&\!\!\!\!\!\!\!\!\!
  +\mbox{(tadpole contributions)},
\label{eq:off-diagonal ghost}
\end{eqnarray}
and
\begin{eqnarray}
&&\!\!\!\!\!\!\!\!\!\!\!\!
D_{\mu\nu}^{ab}(q^2)^{-1}
 =Z_AD_{\mu\nu}^{(0)ab}(q^2)^{-1}
    \nonumber\\
&&\!\!\!\!\!\!\!\!\!
   -Z_A^2
     \int_p
     {\mit\Gamma}_{\rho\mu\xi}^{(0)ac}(r,p)
     D_{\xi\eta}^{cd}(p)
     {\mit\Gamma}_{\sigma\nu\eta}^{db}(p,r)
     D_{\rho\sigma}(r)
    \nonumber\\
&&\!\!\!\!\!\!\!\!\!
  +\mbox{(tadpole contributions)}.
\label{eq:off-diagonal gluon}
\end{eqnarray}
Here the contributions from the two-loop graphs have been omitted.
The full form of SD equations will be given in a separate paper%
\cite{KMS}.
${\mit\Gamma}_\mu^{ac}(p,q)$ is the full vertex function
for the diagonal gluon, off-diagonal ghost and off-diagonal antighost
interaction, while ${\mit\Gamma}_{\mu\rho\sigma}^{ac}(p,q)$
is the full vertex function
for an interaction of the diagonal gluon and two off-diagonal gluons,
and the superscript ``$(0)$'' means a {\it bare} propagator or vertex function.

In the MA gauge, we obtain the Slavnov-Taylor (ST) identities
\begin{equation}
(p-q)^\mu{\mit\Gamma}_\mu^{ab}(p,q)
 ={\mit\Delta}^{ab}(p^2)^{-1}-{\mit\Delta}^{ab}(q^2)^{-1},
\label{eq:STI-C}
\end{equation}
%and
\begin{equation}
(p-q)^\mu{\mit\Gamma}_{\mu\rho\sigma}^{ab}(p,q)
 =D_{\rho\sigma}^{ab}(p^2)^{-1}-D_{\rho\sigma}^{ab}(q^2)^{-1}.
\label{eq:STI-A}
\end{equation}

%%%%%%%%%%%%%%%%%%%%%%%%%%%%%%%%%%%%%%%%%%%%%%%%%%%%%%%%%%
\section{Truncation and Approximations} %
%%%%%%%%%%%%%%%%%%%%%%%%%%%%%%%%%%%%%%%%%%%%%%%%%%%%%%%%%%
In order to solve the SD equations analytically, we employ the
following approximations.
\begin{list}{$\bullet$}{
\setlength{\leftmargin}{12pt}
\setlength{\rightmargin}{0pt}
\setlength{\itemsep}{0pt}
\setlength{\topsep}{3pt}
\setlength{\parsep}{0pt}
}
\item
We neglect the two-loop contributions.

\item
Instead of the full vertex functions, we adopt modified vertex functions
which are compatible with the ST identities.
We adopt approximations for vertex functions as
\begin{equation}
{\mit\Gamma}_\mu^{ab}(p,q)
 \sim{\mit\Gamma}_\mu^{(0)ab}(p,q)\partial_{p^2}\{p^2G^{-1}(p^2)\},
\label{eq:aCC}
\end{equation}
and
\begin{equation}
{\mit\Gamma}_{\mu\rho\sigma}^{ab}(p,q)
 \sim{\mit\Gamma}_{\mu\rho\sigma}^{(0)ab}(p,q)
 \partial_{p^2}\{p^2F_{\rm T}^{-1}(p^2)\}.
\label{eq:aAA}
\end{equation}
Here, we adopt the Feynman gauge
for the off-diagonal gluon
for simplicity, that is, $\alpha=1$ and
$F_{\rm T}(p^2)=F_{\rm L}(p^2)$.
Substituting the bare form factors, which are $G(p^2)=F_{\rm T}(p^2)=1$,
into the right hand side of the ansatz (\ref{eq:aCC})
and (\ref{eq:aAA}),
we obtain the bare vertex functions.
Moreover, these ansatz are compatible with the ST
identities~(\ref{eq:STI-C}) and (\ref{eq:STI-A})
in the limit of $p\rightarrow q$.

\item
In the momentum integration,
we use the Higashijima-Miransky approximation\cite{HM} as
\begin{equation}
F\left((p-q)^2\right)
 =F\left(\max\{p^2,q^2\}\right).
\end{equation}

\end{list}

%%%%%%%%%%%%%%%%%%%%%%%%%%%%%%%%%%%%%%%%%%%%%%%%%%%
\section{Solving the SD equations} %
%%%%%%%%%%%%%%%%%%%%%%%%%%%%%%%%%%%%%%%%%%%%%%%%%%%
Now we adopt the ansatz for the form factors in the IR region:
%\begin{equation}
%F_{\rm d}(p^2)
% =A(p^2)^u+\cdots,
%\label{eq:F_d}
%\end{equation}
%\begin{equation}
%G(p^2)
% =B(p^2)^v+\cdots,
%\label{eq:G}
%\end{equation}
%\begin{equation}
%F_{\rm T}(p^2)
% =C(p^2)^w+\cdots.
%\label{eq:F_T}
%\end{equation}
\begin{equation}
\begin{array}{l}
F_{\rm d}(p^2)
 =A(p^2)^u+\cdots,\\[1mm]
G(p^2)
 =B(p^2)^v+\cdots,\\[1mm]
F_{\rm T}(p^2)
 =C(p^2)^w+\cdots.
\end{array}
\label{eq:IR solutions}
\end{equation}
Substituting the ansatz (\ref{eq:IR solutions}) for the form factors,
and the ansatz~(\ref{eq:aCC}) and (\ref{eq:aAA}) for vertex functions
into the SD equations
(\ref{eq:diagonal gluon}), (\ref{eq:off-diagonal ghost}) and
(\ref{eq:off-diagonal gluon}), and comparing the leading term
in the both sides of each equation, we obtain the following results
for $\alpha=1$.
\begin{list}{}{
\setlength{\leftmargin}{15pt}
\setlength{\rightmargin}{0pt}
\setlength{\itemsep}{0pt}
\setlength{\labelsep}{5pt}
\setlength{\topsep}{3pt}
\setlength{\parsep}{3pt}
\setlength{\listparindent}{0pt}
\setlength{\labelwidth}{10pt}
}
\item[1.]
From eqs.~(\ref{eq:off-diagonal ghost}) and
(\ref{eq:off-diagonal gluon}), we obtain the
relations $v\ge1$ and $w\ge1$.

\item[2a.]
In the case of $v>1$ and $w>1$,
from the eq.~(\ref{eq:diagonal gluon}),
we obtain the relation $u=-\min\{v,w\}$
so that $u$ is less than $-1$.

\item[2b.]
In the case of $v=1$ and $w>1$,
we need redefine the form factor $G(p^2)$ as
$%\begin{equation}
G(p^2)=B_0p^2+B_1(p^2)^{v^\prime}+\cdots
%\label{eq:G'}
$
with $v^\prime>1$
%\end{equation}
since contributions from the leading term of $G(p^2)$
are canceled each other in the ansatz (\ref{eq:aCC}).
Therefore we need the information of next leading term
of the form factor $G(p^2)$.
In this case we obtain the relation $u=-\min\{v^\prime,w\}$
from the eq.~(\ref{eq:diagonal gluon})
so that $u$ is also less than $-1$.

\item[2c.]
Next, we consider the case of $v>1$ and $w=1$.
As well as the above case,
we need redefine the form factor $F_{\rm T}(p^2)$ as
$%\begin{equation}
F_{\rm T}(p^2)=C_0p^2+C_1(p^2)^{w^\prime}+\cdots
%\label{eq:F_T'}
%\end{equation}
$
with $w^\prime>1$
and we obtain the relation $u=-\min\{v,w^\prime\}$
($u<-1$).

\item[2d.]
Similarly, in the case of $v=w=1$,
%making use of the definitions
%(\ref{eq:G'}) and (\ref{eq:F_T'}),
we obtain the relation 
$u=-\min\{v^\prime,w^\prime\}$
($u<-1$).

\end{list}
The results are summarized
in Table~\ref{tbl:Feynman gauge}.
\begin{table}[tb]
\caption{The relation $u$, $v$ and $w$ (and $v^\prime$, $w^\prime$)
in the Feynman gauge ($\alpha=1$).}
\def\arraystretch{1.5}
\begin{tabular}{|c|c|c|}
\hline
      & $w>1$ & $w=1$ \cr
\hline
$v>1$ & $u=-\min\{v,w\}$ & $u=-\min\{v,w^\prime\}$ \cr
\hline
$v=1$ & $u=-\min\{v^\prime,w\}$ & $u=-\min\{v^\prime,w^\prime\}$ \cr
\hline
\end{tabular}
\vspace*{2mm}
\label{tbl:Feynman gauge}
\end{table}

In the gauge other than the Feynman gauge, that is, $\alpha\ne1$,
the calculation and discussion are very tedious.
However, the qualitative results are identical to the above case
except for the following one point.
In this case, even if $w=1$,
there occurs no cancellation as in the above two cases 2c and 2d.
This is because the off-diagonal gluon propagator has
the momentum dependent tensor structure for $\alpha\ne1$,
while it is proportional to $\delta_{\mu\nu}$ for $\alpha=1$.
Therefore, we obtain the relation $u=-1$ in the case of $w=1$.
(See Table~\ref{tbl:not Feynman gauge}.)
\begin{table}[tb]
\caption{The relation $u$, $v$ and $w$ (and $v^\prime$)
in the gauge $\alpha\ne1$.}
\def\arraystretch{1.5}
\begin{tabular}{|c|c|c|}
\hline
      & $w>1$ & $\qquad w=1\qquad$ \cr
\hline
$v>1$ & $u=-\min\{v,w\}$ & $u=-1$ \cr
\hline
$v=1$ & $u=-\min\{v^\prime,w\}$ &$u=-1$ \cr
\hline
\end{tabular}
\vspace*{2mm}
\label{tbl:not Feynman gauge}
\end{table}

%%%%%%%%%%%%%%%%%%%%%%%%%%%%%%%%%%
\section{Conclusion} %
%%%%%%%%%%%%%%%%%%%%%%%%%%%%%%%%%%
%We have derived the SD equations from the $SU(2)$ Yang-Mills theory
%in the MA gauge and solved them analytically in the IR asymptotic region
%under the some approximations.
%We have adopted the ansatz for the vertex functions,
%which are compatible with the ST identity in the IR limit,
%and investigated the qualitative behaviors of propagators.
%Then we have obtained the following results.
In the IR limit, the form factors of each propagator behave as
\begin{equation}
F_{\rm d}(p^2)\sim A(p^2)^u+\cdots
\quad(u\le-1),
\end{equation}
\begin{equation}
G(p^2)\sim B(p^2)^v+\cdots
\quad(v\ge1),
\end{equation}
\begin{equation}
F_{\rm T}(p^2)\sim C(p^2)^w+\cdots
\quad(w\ge1).
\end{equation}
%\begin{equation}
%\begin{array}{l}
%F_{\rm d}(p^2)\sim (p^2)^u+\cdots
%\quad(u<-1),\cr
%G(p^2)\sim (p^2)^v+\cdots
%\quad(v\ge1),\cr
%F_{\rm T}(p^2)\sim (p^2)^w+\cdots
%\quad(w\ge1).
%\end{array}
%\end{equation}
Therefore the solution shows that
the diagonal gluon propagator is enhanced in the IR limit,
while the off-diagonal gluon and off-diagonal ghost propagators are
suppressed in the IR region.
Our results are compatible with a hypothesis of Abelian dominance\cite{EI82}.

%############%
% References %
%############%

\end{document}